\documentclass{emulateapj}

\usepackage{amssymb}

\def\ts     {\thinspace}
\def\kms    {\ts km\ts s$^{-1}$}
\def\etal   {{\rm et\ts al.}}
\def\msol   {$M_{\odot}$}

\def\lprime {K\,\kms\,pc$^2$}

\def\aco    {{\rm CO}($J$=1$\to$0)}

\def\cco    {{\rm CO}($J$=3$\to$2)}

\def\gco    {{\rm CO}($J$=7$\to$6)}

\def\ico    {{\rm CO}($J$=9$\to$8)}

\slugcomment{draft version \today, accepted for publication in the Astrophysical Journal Letters (EVLA Special Issue)}

\shorttitle{\aco\ in $z$$>$2 Quasar Host Galaxies}
\shortauthors{Riechers et al.}

\begin{document}

\title{
\aco\ in $z$$>$2 Quasar Host Galaxies:\\ No Evidence for Extended Molecular Gas Reservoirs}

\author{Dominik A.\ Riechers\altaffilmark{1}, Christopher L.\ Carilli\altaffilmark{2}, Ronald J.\ Maddalena\altaffilmark{3}, Jacqueline Hodge\altaffilmark{4}, Andrew I.\ Harris\altaffilmark{5}, Andrew J.~Baker\altaffilmark{6}, Fabian Walter\altaffilmark{4}, Jeff Wagg\altaffilmark{7}, Paul A.\
Vanden Bout\altaffilmark{8}, Axel Wei\ss\altaffilmark{9}, and Chelsea E. Sharon\altaffilmark{6}}

\altaffiltext{1}{Astronomy Department, California Institute of
  Technology, MC 249-17, 1200 East California Boulevard, Pasadena, CA
  91125, USA; dr@caltech.edu}

\altaffiltext{2}{National Radio Astronomy Observatory, PO Box O, Socorro, NM 87801, USA}

\altaffiltext{3}{National Radio Astronomy Observatory, P.O. Box 2, Green Bank, WV 24944, USA}

\altaffiltext{4}{Max-Planck-Institut f\"ur Astronomie, K\"onigstuhl 17, D-69117 Heidelberg, Germany}

\altaffiltext{5}{Department of Astronomy, University of Maryland, College Park, MD 20742-2421, USA}

\altaffiltext{6}{Department of Physics and Astronomy, Rutgers, the State University of New Jersey, 136 Frelinghuysen Road, Piscataway, NJ 08854-8019, USA}

\altaffiltext{7}{European Southern Observatory, Alonso de C{\'o}rdova 3107, Vitacura, Casilla 19001, Santiago 19, Chile}

\altaffiltext{8}{National Radio Astronomy Observatory, 520 Edgemont
  Road, Charlottesville, VA 22903-2475, USA}

\altaffiltext{9}{Max-Planck-Institut f\"ur Radioastronomie, Auf dem H\"ugel 69, D-53121 Bonn, Germany}


\begin{abstract}

We report the detection of \aco\ emission in the strongly lensed
high-redshift quasars IRAS\,F10214+4724 ($z$=2.286), the Cloverleaf
($z$=2.558), RX\,J0911+0551 ($z$=2.796), SMM\,J04135+10277
($z$=2.846), and MG\,0751+2716 ($z$=3.200), using the Expanded Very
Large Array and the Green Bank Telescope. We report lensing-corrected
\aco\ line luminosities of $L'_{\rm CO} = 0.34-18.4
\times 10^{10}\,$K \kms pc$^2$ and total molecular gas masses of
$M({\rm H_2}) = 0.27-14.7 \times 10^{10}$\,M$_{\odot}$ for the sources
in our sample. Based on CO line ratios relative to previously reported
observations in $J$$\geq$3 rotational transitions and line excitation
modeling, we find that the \aco\ line strengths in our targets are
consistent with single, highly-excited gas components with constant
brightness temperature up to mid-$J$ levels. We thus do not find any
evidence for luminous extended, low excitation, low surface brightness
molecular gas components. These properties are comparable to those
found in $z$$>$4 quasars with existing \aco\ observations.  These
findings stand in contrast to recent \aco\ observations of
$z$$\simeq$2--4 submillimeter galaxies (SMGs), which have lower CO
excitation and show evidence for multiple excitation components,
including some low-excitation gas. These findings are consistent with
the picture that gas-rich quasars and SMGs represent different stages
in the early evolution of massive galaxies.

\end{abstract}


\section{Introduction}

Investigations of the molecular and dusty interstellar medium (ISM) in
high redshift galaxies are of key importance to studies of galaxy
evolution at early cosmic times, as the ISM provides the material that
fuels star formation and stellar mass assembly. It is particularly
interesting to better understand the ISM properties of galaxies that
host both intense star formation and a luminous active galactic
nucleus (AGN), as this enables simultaneous investigations of
supermassive black hole and stellar bulge growth in galaxies. A key
cosmic epoch for such studies is the redshift range
2$\lesssim$$z$$\lesssim$3, where both cosmic star formation and AGN
activity peak, and thus, where most of the growth of stellar and black
hole mass in galaxies occurs (e.g., Magnelli et al.\
\citeyear{mag09}; Richards et al.\ \citeyear{ric06}).

Some of the most remarkable distant AGN-starburst galaxies are
gas-rich, far-infrared luminous quasars at $z$$>$2. To date, the
molecular ISM of 34 high-$z$ quasars has been detected through
emission from rotational lines of CO, 14 of which are gravitationally
lensed (see review by Solomon \& Vanden Bout \citeyear{sv05}, and
Riechers \citeyear{rie11c} for a recent summary). Given the
sensitivity of past observatories, gravitational lensing is key to
probe down to intrinsically fainter (and thus, more common)
systems. One key aspect of these studies is that most detections were
obtained in mid-$J$ (i.e., $J$$\geq$3) CO transitions. Only three of
the 34 quasars were observed in the ground-state \aco\ line (e.g.,
Carilli et al.\ \citeyear{car02}; Riechers et al.\ \citeyear{rie06}).
It thus remains a possibility that many of these studies are biased
toward the highly excited gas that does not necessarily trace the
entire molecular gas reservoir seen in \aco.

To overcome the limitations of previous studies, we have initiated a
systematic study of the \aco\ content of high-$z$ quasars and other
galaxy populations with the Expanded Very Large Array (EVLA; Perley et
al.\ \citeyear{per11}) and the 100\,m Robert C.\ Byrd Green Bank
Telescope (GBT).  In this Letter, we report the detection of \aco\
emission toward five strongly lensed $z$$>$2 quasars, using the EVLA
and the GBT.  We use a concordance, flat $\Lambda$CDM cosmology
throughout, with $H_0$=71\,\kms\,Mpc$^{-1}$, $\Omega_{\rm M}$=0.27,
and $\Omega_{\Lambda}$=0.73 (Spergel \etal\ \citeyear{spe03},
\citeyear{spe07}).

\begin{figure*}
\vspace{-3mm}

\epsscale{1.15}
\plotone{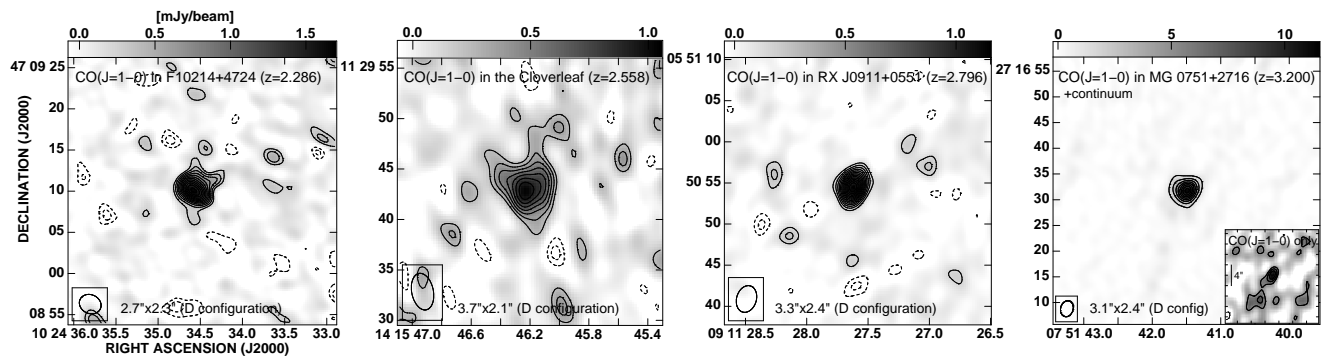}
\vspace{-5mm}

\caption{EVLA \aco\ maps of IRAS\,F10214+4724 ($z$=2.286), the Cloverleaf ($z$=2.558), RX\,J0911+0551 ($z$=2.796), and MG\,0751+2716 ($z$=3.200). The resolution is indicated at the bottom of each panel. For MG\,0751+2716, the inset shows the CO line emission after subtracting the underlying continuum emission. Contours are in steps of 1$\sigma$=125, 115, 75, and 330\,$\mu$Jy\,beam$^{-1}$ for each source, respectively, starting at $\pm$2$\sigma$ (except the CO+continuum map of MG\,0751+2716, where contours are in steps of 4$\sigma$). The maps are averaged over 160, 648, 185, and 306\,\kms\ (18.75, 70, 18.75, and 28\,MHz), respectively.\label{f1}}
%
\end{figure*}

\section{Observations}

\subsection{EVLA}

We observed the \aco\ ($\nu_{\rm rest} = 115.2712$\,GHz) emission line
toward IRAS\,F10214+4724 ($z$=2.286), the Cloverleaf ($z$=2.558),
RX\,J0911+0551 ($z$=2.796), and MG\,0751+2716 ($z$=3.200), using the
EVLA. At these redshifts, all lines are shifted to the Ka band
(0.9\,cm; see Table~\ref{t1} for redshifted line frequencies).
Observations were carried out under good weather conditions in six D
array tracks between 2009 October 26 and December 09, and on 2010 July
17 and 18. This resulted in 5.0, 1.0, 12.0, and 1.0\,hr (2.1, 0.6,
5.1, and 0.6\,hr) total (on-source) observing time for
IRAS\,F10214+4724, the Cloverleaf, RX\,J0911+0551, and MG\,0751+2716,
respectively. For IRAS\,F10214+4724 and RX\,J0911+0551, an additional
0.9 and 2.3\,hr on source were spent on separate continuum settings.
The nearby quasars J0958+4725, J1415+1320, J0909+0121, and J0748+2400
were observed every 3.5 to 7\,minutes for pointing, secondary
amplitude and phase calibration. For primary flux calibration, the
standard calibrators 3C286 and 3C147 were observed, leading to a
calibration that is accurate within $\sim$10\%.

Observations for the Cloverleaf and MG\,0751+2716 were carried out
with the WIDAR correlator, using two intermediate frequencies (IFs) of
128\,MHz (dual polarization) each at 2\,MHz resolution. For the
Cloverleaf, the two IFs were overlapped by two channels, centered on
the CO line, yielding 252\,MHz contiguous bandwidth. For
MG\,0751+2716, one IF was centered on the CO line, and the second IF
was centered on the continuum at 32.046\,GHz.

Observations for IRAS\,F10214+4724 and RX\,J0911+0551 (which have
narrow CO lines) were carried out with the previous generation
correlator, with two 21.875\,MHz (dual polarization) IFs at 3.125\,MHz
resolution. For IRAS\,F10214+4724, both IFs were centered on the CO
line, yielding 43.75\,MHz contiguous bandwidth. For RX\,J0911+0551,
one IF was centered on the CO line, and the second IF was centered on
the continuum at 34.3173\,GHz. For these two sources, one third of the
on-source time was spent to observe a second frequency setting with
two 50\,MHz continuum IFs offset by $\pm$150\,MHz from the CO lines,
yielding more sensitive constraints on the continuum emission.

For data reduction and analysis, the AIPS package was used.  All data
were mapped using `natural' weighting.  The data result in final rms
noise levels of 125, 115, 75, and 330\,$\mu$Jy\,beam$^{-1}$ over 160,
648, 185, and 306\,\kms\ (18.75, 70, 18.75, and 28\,MHz) widths for
IRAS\,F10214+4724, the Cloverleaf, RX\,J0911+0551, and MG\,0751+2716,
respectively. Maps of the velocity-integrated CO $J$=1$\to$0 line
emission yield synthesized clean beam sizes of 2.7$''$$\times$2.3$''$,
3.7$''$$\times$2.1$''$, 3.3$''$$\times$2.4$''$, and
3.1$''$$\times$2.4$''$.

\begin{deluxetable*}{lcccccccc}
\vspace{-7mm}
\tabletypesize{\scriptsize}
\tablecaption{Observed \aco\ line parameters. 
\label{t1}}
\tablehead{
Source          & $z_{\rm CO}$        & $\mu_{\rm L}$ & $\nu_{\rm obs}$ & $S_{\nu}$       & $\Delta V_{\rm FWHM}$ & $I_{\rm CO}$ & $L'_{\rm CO(1-0)}$\tablenotemark{a} & Telescope \\
                &                     &               & [GHz]           & [mJy]           & [\kms]                & [Jy \kms]    & [10$^{9}$\,K\,\kms\,pc$^2$]     &
}
\startdata
F10214+4724     & 2.2853 $\pm$ 0.0001 & 17  & 35.0795         & 2.42 $\pm$ 0.37 & 169 $\pm$ 39          & 0.434 $\pm$ 0.047 & & EVLA \\
                & 2.2856 $\pm$ 0.0001 &     &                 & 1.73 $\pm$ 0.22 & 184 $\pm$ 29          & 0.337 $\pm$ 0.045 & & GBT/S \\
                & 2.2854 $\pm$ 0.0001 &     &                 &                 &                       & 0.383 $\pm$ 0.032 & 5.77$\pm$0.49 & Combined \\
\tableline
Cloverleaf      & 2.5564 $\pm$ 0.0004 & 11  & 32.3992         & 2.78 $\pm$ 0.40 & 468 $\pm$ 94          & 1.378 $\pm$ 0.250 & & EVLA \\
                & 2.5578 $\pm$ 0.0001 &     &                 & 2.72 $\pm$ 0.12 & 470 $\pm$ 27          & 1.358 $\pm$ 0.065 & & GBT/S \\
                & 2.5574 $\pm$ 0.0001 &     &                 & 3.14 $\pm$ 0.11 & 422 $\pm$ 19          & 1.406 $\pm$ 0.053 & & GBT/Z \\
                & 2.5575 $\pm$ 0.0001 &     &                 &                 &                       & 1.387 $\pm$ 0.040 & 39.3$\pm$1.1 & Combined \\
\tableline
J0911+0551      & 2.7961 $\pm$ 0.0001 & 22  & 30.3665         & 1.75 $\pm$ 0.22 & 111 $\pm$ 19          & 0.205 $\pm$ 0.029 & 3.39$\pm$0.48 & EVLA \\
\tableline
J04135+10277    & 2.8470 $\pm$ 0.0004 & 1.3 & 29.9717         & 1.20 $\pm$ 0.15 & 505 $\pm$ 75          & 0.644 $\pm$ 0.082 & 184$\pm$23  & GBT/S \\
\tableline
MG\,0751+2716   & 3.1984 $\pm$ 0.0006 & 16  & 27.4455         & 1.61 $\pm$ 0.34 & 290 $\pm$ 62          & 0.494 $\pm$ 0.105 & & EVLA \\
                & 3.1995 $\pm$ 0.0004 &     &                 & 1.47 $\pm$ 0.24 & 354 $\pm$ 72          & 0.550 $\pm$ 0.095 & & GBT/S \\
                & 3.1990 $\pm$ 0.0003 &     &                 &                 &                       & 0.525 $\pm$ 0.070 & 14.9$\pm$2.0 & Combined \\
\vspace{-3mm}
\enddata
\tablenotetext{a}{Corrected for magnification due to gravitational lensing.} 
\tablecomments{$\mu_{\rm L}$:\ lensing magnification factor (see Riechers \citeyear{rie11c}, and references therein). GBT/S:\ Measured with the GBT digital Spectrometer backend. GBT/Z:\ Measured with the GBT wideband Zpectrometer backend.
}
\end{deluxetable*}

\begin{figure*}
\vspace{-5mm}
\epsscale{1.15}
\plotone{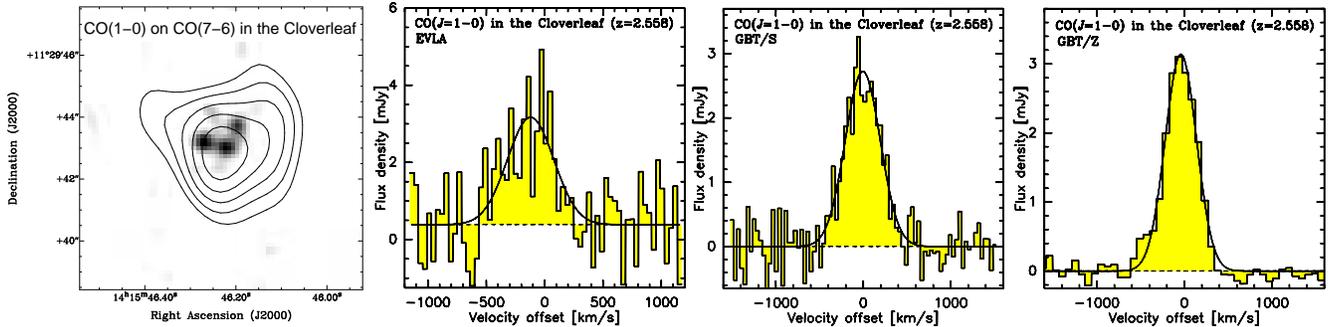}
\vspace{-2mm}

\caption{Overlay of \aco\ and \gco\ emission (left; CO $J$=7$\to$6 map from Alloin et al.\ \citeyear{all97}), and independent EVLA, GBT/Spectrometer, and GBT/Zpectrometer \aco\ spectra (middle left to right) in the Cloverleaf ($z$=2.558). The spectra (histograms) are shown at 4, 3.906, and 8\,MHz resolution (37, 36, and 74\,\kms ). The solid curves indicate Gaussian fits to the spectra. The GBT/Z channels are not statistically independent on scales below 20\,MHz (2.5 channels).\label{f2}}
%
\end{figure*}

\subsection{GBT}

We observed the \aco\ emission line toward IRAS\,F10214+4724, the
Cloverleaf, SMM\,J04135+10277 ($z$=2.846), and MG\,0751+2716, using
the facility Ka band receiver on the GBT with the digital Spectrometer
backend (GBT/S) and the Zpectrometer analog lag cross-correlation
spectrometer backend (GBT/Z; instantaneously covering 25.6--36.1\,GHz;
Harris et al.\ \citeyear{har07}; see Table~\ref{t1}). This yields a
typical beam size of $\sim$23$''$. GBT/S observations were carried out
under acceptable to good weather conditions during 14 sessions between
2007 October 10 and 2008 October 12, yielding typical system
temperatures of $T_{\rm sys}$=42--57\,K (with higher $T_{\rm sys}$
toward higher observing frequencies due to the atmosphere).  This
resulted in 4.3, 3.7, 2.5, and 8.7\,hr on-source observing time for
IRAS\,F10214+4724, the Cloverleaf, SMM\,J04135+10277, and
MG\,0751+2716, respectively.  In addition, the Cloverleaf was observed
with the GBT/Z for another 6 sessions between 2008 March 02 and 2009
March 12.  This resulted in 14.7\,hr of observing time, about half of
which was spent on the Cloverleaf. The GBT/S was configured with
800\,MHz bandwidth, yielding a spectral resolution of 391\,kHz. The
GBT/Z samples its 10.5\,GHz bandwidth with 8\,MHz spectral channels.
Its instrumental spectral response is nearly a sinc function with an
FWHM of 20\,MHz, i.e., individual 8\,MHz spectral channels are not
statistically independent. However, the line width correction for the
instrumental response for spectral lines with intrinsic Gaussian FWHM
of $>$30\,MHz ($\sim$300\,\kms ) is minor.

Subreflector beam switching was used every 10\,s for all observations
(GBT/S and GBT/Z) to observe the sources alternately with the
receiver's two beams, with the off-source beam monitoring the sky
background in parallel. To remove continuum fluxes and
atmospheric/instrumental effects, low-order polynomials were fitted to
the the spectral baselines of the calibrated, time-averaged GBT/S
observations. For the GBT/Z observations, a nearby second, faint
source was observed with the same subreflector switching pattern,
alternating between targets with 8\,min cycles. Residual structure
from optical beam imbalance in the ``source--sky'' difference spectra
of the two targets was then removed by differencing the resulting
spectra of both sources (the second source was not detected).  This
strategy yields a flat baseline (offset from zero flux by the
difference of the two source's continua) without standard polynomial
baseline removal.  For all observations, several nearby quasars were
targeted regularly to monitor telescope pointing and gain
stability. Passband gains and absolute fluxes for all observations
were determined from spectra of 3C286 and 3C147, yielding 10\%--15\%
calibration accuracy.

All data were processed with GBTIDL (Marganian et al.\
\citeyear{mag06}), using standard recipes and/or the Zpectrometer's
data reduction pipeline (see Harris et al.\ \citeyear{har10} for GBT/Z
details). After processing, the GBT/S data were re-binned to
3.906\,MHz (33--43\,\kms ) for further analysis.

\begin{figure*}
\vspace{-2mm}
\epsscale{1.15}
\plotone{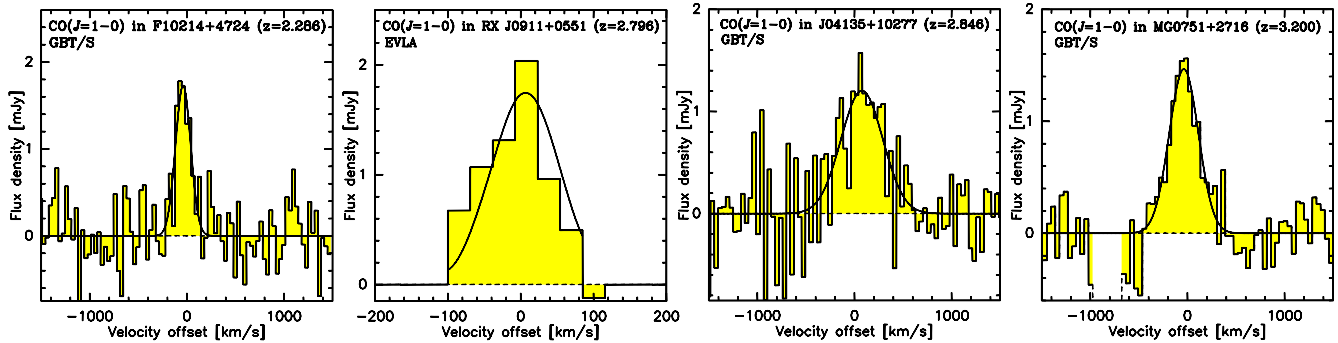}

\caption{EVLA and GBT/Spectrometer spectra of \aco\ emission toward IRAS\,F10214+4724 ($z$=2.286), RX\,J0911+0551 ($z$=2.796), SMM\,J04135+10277
($z$=2.846), and MG\,0751+2716 ($z$=3.200). The spectra (histograms) are shown at 3.906\,MHz resolution (33--43\,\kms ), except RX\,J0911+0551, which is
shown at 3.125\,MHz (31\,\kms ) resolution. The solid curves indicate Gaussian fits to the spectra. The blanked-out region in the spectrum of MG\,0751+2716 masks a resonance from the GBT Ka band receiver's feeds. \label{f3}}
%
\end{figure*}

\section{Results}

We have detected strong \aco\ emission toward IRAS\,F10214+4724, the
Cloverleaf, RX\,J0911+0551, SMM\,J04135+10277, and MG\,0751+2716. We
spatially resolve the \aco\ emission in IRAS\,F10214+4724 and the
Cloverleaf, but do not resolve the \aco\ emission in RX\,J0911+0551
and MG\,0751+2716 (Fig.~\ref{f1}). The measured sizes and upper limits
are consistent with the lens configurations of these targets if the CO
reservoirs have intrinsic sizes of a few kpc. Accounting for beam
convolution, the gas distribution in the Cloverleaf is consistent with
that observed in higher-$J$ lines (e.g., Alloin et al.\
\citeyear{all97}), but may be somewhat more extended (Fig.~\ref{f2},
left).  We detect rest-frame 2.6\,mm continuum emission toward the
Cloverleaf and MG\,0751+2716 at 386$\pm$107\,$\mu$Jy and
17.4$\pm$0.7\,mJy strength. We also detect rest-frame 2.2\,mm
continuum emission toward MG\,0751+2716 at 13.9$\pm$0.6\,mJy
strength. We do not detect rest-frame 2.6\,mm continuum emission
toward IRAS\,F10214+4724 and RX\,J0911+0551 down to 3$\sigma$ limits
of 270 and 170\,$\mu$Jy. The continuum emission in MG\,0751+2716 is
consistent with non-thermal emission from its radio-loud AGN (e.g.,
Leh\'ar et al.\ \citeyear{leh97}; Riechers et al.\
\citeyear{rie06b}). The continuum detection and limits for the other
targets are consistent with thermal and non-thermal emission
associated with star formation in their host galaxies (e.g., Wei\ss\
et al.\ \citeyear{wei03}; Ao et al.\ \citeyear{ao08}; Wu et al.\
\citeyear{wu09}).

The \aco\ line profiles of all sources were fitted with
single-component Gaussians, which yields good fits to all data
(Figs.~\ref{f2} and \ref{f3}). The three independent measurements of
the \aco\ line in the Cloverleaf (EVLA, GBT/S, and GBT/Z) yield
consistent line profiles and intensities within the relative
uncertainties, suggesting that the interferometric observations
recover the full flux, and that the calibration of the single-dish
observations is reliable to high accuracy. Due to the limited
bandwidth of the observations of RX\,J0911+0551, the line is only
marginally covered in velocity. However, the fitted line FWHM of
111$\pm$19\,\kms\ is in excellent agreement with the $\sim$110\,\kms\
measured in the \cco\ to \ico\ lines (A.~Wei\ss\ et al., in
prep.).\footnote{These measurements revise the \cco\ line flux and
width measured by Hainline et al.\ \citeyear{hai04} down by $\sim$40\%
and by a factor of $\sim$3, respectively.}  The GBT/S measurement of
the \aco\ line in MG\,0751+2716 shows an instrumental resonance
feature $\sim$800\,\kms\ bluewards from the line center, which is
excluded from the spectral baseline fit, and does not appear to affect
the line flux measurement. The \aco\ line peak flux densities, widths,
integrated intensities, and centroid redshifts for all sources and
measurements are detailed in Table~\ref{t1}. Two or three independent
\aco\ measurements were obtained for IRAS\,F10214+4724, the
Cloverleaf, and MG\,0751+2716. The median CO redshifts and line
intensities for these sources are given in Table~\ref{t1}, and are
adopted for all further analysis. All five targets are gravitationally
lensed. Based on the \aco\ line intensities and lensing magnification
factors from the literature (see Riechers \citeyear{rie11c}, and
references therein), we derive lensing-corrected \aco\ line
luminosities of $L'_{\rm CO(1-0)} = 0.34-18.4\times 10^{10}\,$\lprime\
(Tab.~\ref{t1}).

Our \aco\ fluxes imply \cco/\aco\ line brightness temperature
ratios\footnote{The error bars for $r_{31}$ are derived from the
statistical uncertainties of the \aco\ and \cco\ measurements.}  of
$r_{31}$=$T_{\rm b}^{\rm CO(3-2)}$/$T_{\rm b}^{\rm
CO(1-0)}$=1.00$\pm$0.10, 1.06$\pm$0.03, $\sim$0.95, 0.93$\pm$0.25, and
0.97$\pm$0.17 for IRAS\,F10214+4724, the Cloverleaf, RX\,J0911+0551,
SMM\,J04135+10277, and MG\,0751+2716, respectively (CO $J$=3$\to$2
measurements are from Ao et al.\ \citeyear{ao08}, Wei\ss\ et al.\
\citeyear{wei03}, A.~Wei\ss\ et al., in prep., Hainline et al.\
\citeyear{hai04}, and Alloin et al.\ \citeyear{all07},
respectively). These values are fully consistent with thermalized gas
excitation (i.e., $r_{31}$=1) in all targets.

\section{Analysis}

\subsection{Line Excitation Modeling}

The two best-studied targets in our sample are IRAS\,F10214+4724 and
the Cloverleaf (e.g., Ao et al.\ \citeyear{ao08}; Barvainis et al.\
\citeyear{bar97}; Wei\ss\ et al.\ \citeyear{wei03}; Bradford et al.\ 
\citeyear{bra09}). Based on \cco\ to \ico\ observations in the
literature and our \aco\ detections, we can constrain the line
radiative transfer through Large Velocity Gradient (LVG) models,
treating the gas kinetic temperature and density as free parameters.
For all calculations, the H$_2$ ortho--to--para ratio was fixed to
3:1, the cosmic microwave background temperature was fixed to 8.95 and
9.69\,K (at $z$=2.286 and 2.558), and the Flower (\citeyear{flo01}) CO
collision rates were used. For consistency with previous modeling of
both sources (Ao et al.\ \citeyear{ao08}; Riechers et al.\
\citeyear{rie11b}), we adopted a CO abundance per velocity gradient of
[CO]/(${\rm d}v/{\rm d}r) = 1 \times 10^{-5}\,{\rm pc}\,$(\kms)$^{-1}$
(e.g., Wei\ss\ \etal\ \citeyear{wei05b}, \citeyear{wei07}; Riechers
\etal\ \citeyear{rie06}).

Observations of both targets are fit very well by single-component
models (except the CO $J$=5$\to$4 flux in the Cloverleaf, which is
likely too low due to calibration issues related to the restricted
bandwidth of these early observations; Barvainis et al.\
\citeyear{bar97}; see also A.~Wei\ss\ \etal, in prep.). For
IRAS\,F10214+4724, we fit a representative model with a kinetic gas
temperature of $T_{\rm kin}$=60\,K and a gas density of $\rho_{\rm
gas}$=10$^{3.8}$\,cm$^{-3}$, yielding a moderate optical depth of
$\tau_{\rm CO(1-0)}$=1.6 (Fig.~\ref{f4}a and c). For the Cloverleaf,
we fit a representative model with $T_{\rm kin}$=50\,K and $\rho_{\rm
gas}$=10$^{4.5}$\,cm$^{-3}$, yielding a high optical depth of
$\tau_{\rm CO(1-0)}$=8.9 (Fig.~\ref{f4}b and d). While not formally
excluding the presence of some colder gas (given the remaining
uncertainties), these models are consistent with previous fits based
on the CO $J$$\geq$3 transitions alone (Ao et al.\ \citeyear{ao08};
Bradford et al.\ \citeyear{bra09}; Riechers et al.\
\citeyear{rie11b}). They are also consistent with what is found for
$z$$>$4 quasars observed in \aco\ emission (Riechers et al.\
\citeyear{rie06}; Wei\ss\ et al.\ \citeyear{wei07}). This finding is
in agreement with the \aco\ emission in high-$z$ quasars being
associated with optically thick emission from the highly excited
molecular gas detected in high-$J$ CO transitions, without any
evidence for significant additional low-excitation gas components.

\begin{figure*}
\vspace{-2mm}
\epsscale{1.15}
\plotone{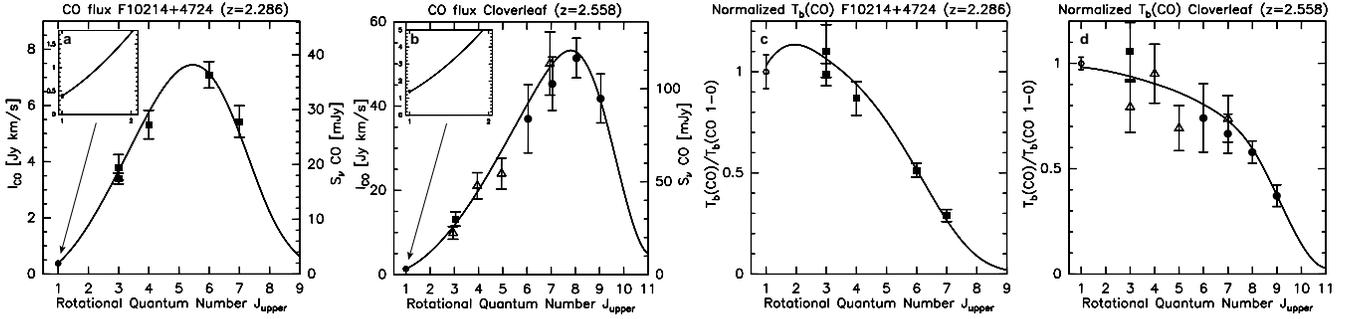}
\vspace{-1mm}

\caption{
CO excitation ladders (points) and LVG models (lines) for
IRAS\,F10214+4724 (panels a and c) and the Cloverleaf (panels b and
d), scaled by CO intensity (left two panels) and normalized brightness
temperature $T_{\rm b}$ (right two panels). The insets show zoomed-in
versions of the model fits close to the
\aco\ line.  The CO data for IRAS\,F10214+4724 (circle:\ this work;
squares:\ Ao et al.\ \citeyear{ao08}) and the Cloverleaf (circle:\
this work; triangles:\ Barvainis et al.\ \citeyear{bar97}; filled
square:\ Wei\ss\ et al.\ \citeyear{wei03}; filled circles:\ Bradford
et al.\ \citeyear{bra09}; error bars of literature data have been
conservatively extended to include 10\%--15\% calibration
uncertainties) are fit well by single, highly excited gas components
with kinetic temperatures of $T_{\rm kin}$=60 and 50\,K and gas
densities of $\rho_{\rm gas}$=10$^{3.8}$ and 10$^{4.5}$\,cm$^{-3}$,
respectively (see also Ao et al.\ \citeyear{ao08}; Riechers et al.\
\citeyear{rie11b}). \label{f4}}
%
\end{figure*}

\subsection{Molecular Gas Masses}

The \aco\ emission in our targets appears to be associated with highly
excited gas that has physical properties consistent with those found
in the nuclei of nearby ultra-luminous infrared galaxies (ULIRGs). We
thus calculate their gas masses assuming a ULIRG-like conversion
factor of $\alpha_{\rm CO}$=0.8\,\msol\,(\lprime )$^{-1}$ from
$L'_{\rm CO(1-0)}$ to $M_{\rm gas}$ (Downes \& Solomon
\citeyear{ds98}). This yields lensing-corrected molecular gas masses
of $M_{\rm gas}$=0.46, 3.1, 0.27, 14.7, and
1.2$\times$10$^{10}$\,\msol\ for IRAS\,F10214+4724, the Cloverleaf,
RX\,J0911+0551, SMM\,J04135+10277, and MG\,0751+2716,
respectively. SMM\,J04135+10277 is now revealed as the high-$z$ quasar
with the highest $M_{\rm gas}$ currently known. The gas masses of all
systems lie within the range found for other high-$z$ quasars (mostly
determined based on observations of $J$$\geq$3 CO transitions), with
IRAS\,F10214+4724 and RX\,J0911+0551 being situated at the low end of
the observed range (which can currently only be investigated with the
aid of strong gravitational lensing; e.g., Riechers
\citeyear{rie11c}).

\section{Discussion and Conclusions}

We have detected strong \aco\ emission toward IRAS\,F10214+4724
($z$=2.286), the Cloverleaf ($z$=2.558), RX\,J0911+0551 ($z$=2.796),
SMM\,J04135+10277 ($z$=2.846), and MG\,0751+2716 ($z$=3.200). Our EVLA
observations have spatially resolved the emission toward
IRAS\,F10214+4724 and the Cloverleaf (Fig.~\ref{f1}). We find line
brightness temperatures consistent with those measured in mid-$J$ CO
lines. Excitation modeling indicates that the \aco\ emission is
associated with the warm, highly excited gas in the star-forming
regions in the host galaxies that is also seen in the higher-$J$ CO
lines. This result suggests that mid-$J$ CO lines are good indicators
of the total amount of molecular gas in gas-rich high redshift quasar
host galaxies, consistent with previous findings based on a smaller
sample of $z$$\gtrsim$4 quasars (Riechers et al.\ \citeyear{rie06}),
and extending these studies to the peak epoch of cosmic star formation
and AGN activity.  In contrast, recent studies of \aco\ emission in
$z$$>$2 SMGs suggest that many of these galaxies have substantial
amounts of low-excitation gas (e.g., Hainline et al.\
\citeyear{hai06}; Carilli et al.\ \citeyear{car10}; Harris et al.\
\citeyear{har10}; Ivison et al.\ \citeyear{ivi10}, \citeyear{ivi11};
Riechers et al.\ \citeyear{rie10}, \citeyear{rie11},
\citeyear{rie11d}).  Such a difference is in agreement with the
picture that gas-rich quasars and SMGs represent different stages in
the early evolution of massive galaxies. Such an observational finding
would be consistent with a high redshift analogue of the ULIRG-quasar
transition scenario proposed by Sanders et al.\ (\citeyear{san88}).

In this picture, it may be expected to find some sources with gas
properties that overlap with both populations, i.e., SMGs with
molecular gas properties closer to those of the quasars studied
here. The recently identified $z$=2.957 SMG HLSW-01 may be an example
of such ``overlap'' sources. Most of its gas properties are consistent
with those of ``typical'' SMGs (i.e., its gas mass, gas mass fraction,
gas depletion timescale, star formation efficiency, specific star
formation rate, and dynamical structure; Riechers et al.\
\citeyear{rie11e}), but it shows a high $r_{31}$=0.95$\pm$0.10 (i.e.,
consistent with 1), and an overall high CO excitation (Riechers et
al.\ \citeyear{rie11e}; Scott et al.\ \citeyear{sco11}). Its high
radio luminosity, dust temperature, and CO excitation may suggest the
presence of a luminous, perhaps obscured AGN (Conley et al.\
\citeyear{con11}; Riechers et al.\ \citeyear{rie11e}; Scott et al.\ 
\citeyear{sco11}). However, based on CO studies at high $z$ to date, 
systems like HLSW-01 appear to be rare among SMGs.

These results demonstrate that observations of low-order CO
transitions in high redshift galaxies with the EVLA will be of
enormous utility to establish the context for studies of highly
excited CO emission with the Atacama Large (sub)Millimeter Array
(ALMA). Complementary studies with the EVLA and ALMA will be key to
distinguishing different high-$z$ galaxy populations based on the
physical properties of the molecular gas in their star-forming
environments.

\acknowledgments 

This letter is dedicated to the memory of Phil Solomon. His
contributions to the field were a true inspiration, and we gratefully
acknowledge his wisdom in the early discussions of the GBT program.
We thank the referee, Dr.~Philip Maloney, for a helpful report.  We
thank Charles C.\ Figura for assistance in the development of data
reduction techniques related to the GBT Spectrometer observations. We
thank Christian Henkel for the original version of the LVG code.  DR
acknowledges support from NASA through a Spitzer Space Telescope
grant.  AJB acknowledges support from NSF grant AST-0708653 to Rutgers
University. The National Radio Astronomy Observatory is a facility of
the National Science Foundation operated under cooperative agreement
by Associated Universities, Inc.

\end{document}